\documentstyle[12pt]{article}
\newcommand{\la}{\langle}
\newcommand{\ra}{\rangle}

\begin{document}
\title{Analysis of Cumulant Moments in High Energy Hadron-Hadron 
Collisions by Truncated Multiplicity Distributions}
\author{Noriaki Nakajima, Minoru Biyajima, 
         and Naomichi Suzuki$^{1}$ \\
 ${}^{}$Department of Physics, Shinshu University, Matsumoto 390, 
Japan \\
 ${}^{1}$Matsusho Gakuen Junior College, Matsumoto 390-12, Japan }  
\date{}
\maketitle
\begin{abstract}
Oscillatory behavior of cumulant moments obtained from the 
experimental data in $pp$ collisions and $\bar{p}p$ collisions are 
analyzed by the modified negative binomial distribution (MNBD) and 
the negative binomial distribution (NBD).  Both distributions well 
describe the cumulant moments obtained from the data.  This fact 
shows sharp contrast to the result in $e^+e^-$ collisions, which is 
described by the the MNBD much better than by the NBD.
\end{abstract}

\section{Introduction}

 Recently a prediction is made by the QCD calculations that the 
cumulant moment of multiplicity distribution possesses an oscillatorye
behavior\cite{drem94a}.  Furthermore, analysis of the cumulant 
moments in hadron-hadron $hh$ and $e^+e^-$ collisions shows that the 
$j$-th order normalized cumulant moment of observed charged 
multiplicity distributions oscillates irregularly around the zero 
with increasing the rank $j$ \cite{drem94b}.  At present calculated 
results by the QCD explain the behavior of the data only 
qualitatively. 

It is well known that the cumulant moment of negative binomial 
distribution (NBD) calculated from the multiplicity generating 
function is positive and decreases monotonously as the rank 
increases.  However, it is shown in Ref.\cite{ugoc95} that 
oscillatory behavior of cumulant moment appears if it is calculated 
from the truncated NBD.
 The possibility is pointed out that the cumulant moments of the 
modified negative binomial distribution (MNBD) calculated from the 
generating function change sign alternatively as the rank of the 
cumulant moments increases \cite{suzu96}.  
Cumulant moments of negatively charged particles and charged 
particles obtained from the data in $e^+e^-$ collisions have been 
analyzed by the MNBD \cite{suzu96}. The result shows that the 
cumulant moments of the negatively charged particles obtained from 
the data show rather regular oscillation around zero; those with evenh
rank are smaller than two adjacent moments at least up to the 6th 
rank.
The cumulant moments of the negatively charged and charged particles 
obtained from the data are well described by the MNBD, if it is 
truncated at the maximum of the observed negatively charged 
multiplicity.  However, in $e^+e^-$ collisions, oscillation of 
cumulant moments calculated from the truncated NBD is very weak 
comparing with that obtained from the data. 

In this paper, we analyze the cumulant moments of charged particles 
observed in $pp$ and $\bar{p}p$ collisions, by the MNBD and the NBD.

The j-th order normalized cumulant $K_j$ of charged particles is 
expressed by the normalized factorial moments $F_l\quad(l=
1,2,\cdots)$ of charged particles as,
 \begin{eqnarray}
        K_1 &=& F_1,      \nonumber  \\
        K_j &=& F_j + \sum_{m=1}^{j-1} {}_{j-1}C_{m-1}\,F_{j-m}\,K_m,,
\qquad
                j=2,3,\cdots,    \label{eq:kq1}
 \end{eqnarray}
where
 \begin{eqnarray*}
  F_j = <n_{\rm ch}(n_{\rm ch}-1)\cdots(n_{\rm ch}-j+1)> 
/<n_{\rm ch}>^j.
 \end{eqnarray*}
The $H_j$ moment is defined by
 \begin{eqnarray}
      H_j = K_j/F_j.       \label{eq:kq2}
 \end{eqnarray}

\section{A Stochastic Process }
   In \cite{suzu91}, a birth process with an immigration is taken forc
a model of particle productions,
  \begin{eqnarray}
      \frac{\partial P(n;t)}{\partial t} =
           - \lambda_{0} P(n;t) + \lambda_0 P(n-1;t)  \nonumber \\
           - \lambda_{2} nP(n;t) + \lambda_2 (n-1)P(n-1;t),   \label
{eq:kq3}
  \end{eqnarray}
where $\lambda_0$ denotes an immigration rate, and $\lambda_2$ a 
birth rate. 
An initial condition of Eq.(\ref{eq:kq3}) is taken as a binomial 
distribution,
  \begin{equation}
    P(n;t=0) = {}_N\,C_n\,\alpha^n\,\beta^{N-n},  
      \quad \beta=1-\alpha.   \label{eq:kq4}
 \end{equation}
 The generating function $\Pi (z)$ of the multiplicity distribution 
$P(n)$ is defined by
 \begin{equation}
         \Pi(z) = \sum_{n=0} P(n)z^n.          \label{eq:kq5}
 \end{equation}
Then, we have the generating function for the distribution $P(n;t=
T)$,
 \begin{eqnarray}
      \Pi(z;t=T)= \frac{1}{[1-p(z-1)]^k}
      \left( \frac{1-[p-\alpha(1+p)](z-1)}{1-p(z-1)} \right)^N, 
\label{eq:kq6}
 \end{eqnarray}
where
 \begin{eqnarray*}
     p=\exp[\lambda_2T]-1,   \qquad   k=\lambda_0/\lambda_2.
 \end{eqnarray*}
If the branching process is governed by neutral particles, and those 
particles finally decay into charged pairs with a probability $a$ or 
neutral particles with $(1-a)$. Then, from Eq.(\ref{eq:kq6}), we havey
the generating function for charged particles, 
  \[  G(w) = \Pi(a(w^2-1)+1;t=T),   \]
where the variable $z$ in Eq.(\ref{eq:kq6}) is replaced by $aw^2+(1-
a)$. The generating function $\Pi(z)$ for negatively charged 
particles is therefore written as
 \begin{eqnarray}
    \Pi(z)&=& \left[ 1 - r_1(z-1) \right]^N
           \left[ 1 - r_2(z-1) \right]^{-N-k},    \nonumber \\
      r_1 &=& a\left( p - \alpha(1+p) \right),   \quad
      r_2 = ap.          \label{eq:kq7}   
 \end{eqnarray}
It should be noted that in Eq.(\ref{eq:kq7}), $N$ is a positive 
integer, $r_1$ is real ($r_1<0$ or $r_1 \ge 0$), and $r_2>0$.

The probability distribution is obtained from $\Pi(z)$ as
 \begin{eqnarray}
    P(0) &=& \Pi (0) = \frac{\left(1+r_1 \right)^N}{\left(1+r_2 
\right)^{N+k}},
                                \nonumber \\
    P(n) &=& \frac{1}{n!}
             \left. \frac{\partial^n \Pi(z)}{\partial z^n} \right|_
{z=0} 
                                \nonumber \\
         &=& \frac{1}{n!} \left( \frac{r_1}{r_2} \right)^N   
              \sum_{j=0}^N {}_N C_j \frac{\Gamma(k+n+j)}{\Gamma(k+j)}
              \left(\frac{r_2-r_1}{r_1} \right)^j 
             \frac{ r_2 ^n }{ \left( 1+r_2 \right)^
{n+k+j} },
                \quad  n=1,2, \cdots.    \label{eq:kq8}
 \end{eqnarray}
If $k=0$, the summation on the right hand side of Eq.(\ref{eq:kq8}) 
runs from $j=1$ up to $j=N$, and it is called the MNBD \cite{suzu91}, 
\cite{chli90}.  The second paper in \cite{chli90} corresponds to the 
case $\alpha =1$, which leads to $r_1<0$. If $N=0$, Eq.(\ref{eq:kq8})a
is reduced to the NBD.  

If $k=0$, parameters $r_1$ and $r_2$ are expressed by the average 
multiplicity $<n>$ and $C_2$ moment of negatively charged particles 
respectively as,
 \begin{eqnarray*}
    r_1 &=& \frac{1}{2} 
        \left( C_2 - 1 - \frac{1}{<n>} - \frac{1}{N} \right)<n>,    \\
    r_2 &=& \frac{1}{2} 
       \left( C_2 - 1 - \frac{1}{<n>} + \frac{1}{N} \right)<n>.
 \end{eqnarray*}
If $N=0$, parameter $r_1$ is not contained in Eq.(\ref{eq:kq8}), and 
$r_2$ is given by
  \[   r_2 = \frac{<n>}{k}.   \]

\section{Analysis of the experimental data}

  The cumulant moments of charged particles at the ISR in $pp$ 
collisions \cite{brea84}, and  of the UA5 collaboration in $ \bar{p}pn
$ collisions \cite{anso87} are analyzed by the MNBD and NBD in this 
paper. 
 Equation (\ref{eq:kq8}) is applied to the multiplicity distribution 
of negatively charged particles. The parameters used in the analysis 
are shown in Table I.  Those are determined by the minimum chi-square 
( $\chi^2_{\rm min}$ ) fit to the observed multiplicity distributionsm
of negatively charged particles. As is seen from the Table I, the 
$\chi^2_{\rm min}$ values of the MNBD fit are smaller than those of 
the NBD fit in both energy regions.

Factorial moments of charged particles are calculated as
 \begin{eqnarray}
   f_j^{{\rm ch}} &=& <n_{\rm ch}(n_{\rm ch}-1)\cdots(n_{\rm ch}-j+
1)>
                                    \nonumber \\ 
       &=& \sum_{n}^{n_{\rm max}}\, (2n+n_0)(2n+n_0-1)\cdots(2n+n_0-
j+1)\,P(n),
                     \quad j=1,2,\cdots,   \label{eq:kq9}
 \end{eqnarray}
where $P(n)$ is given by Eq.(\ref{eq:kq8}).
In Eq.(\ref{eq:kq9}), $n_{\rm max}$ denotes the maximum of the 
observed negatively charged multiplicity, and $n_0$ is taken as $n_0=
2$ for $pp$ collisions, and $n_0=0$ for $\bar{p}p$ collisions.

Then, cumulant moments of charged particles are calculated from Eqs.
(\ref{eq:kq1}), (\ref{eq:kq2}) and (\ref{eq:kq9}) with the parameters{
shown in Table I.  Those are compared with the cumulant moments of 
charged particles at the ISR in Figs.1a, b, c and d, and those for 
the UA5 Collaboration from $200$ GeV to $900$ GeV in Figs.2a, b and 
c.  Both the MNBD and the NBD well describe the behavior of the data.

%
\section{Concluding remarks}

  The cumulant moments of observed multiplicity distributions of 
charged particles in $pp$ and $\bar{p}p$ collisions are analyzed by 
the MNBD and the NBD.  The cumulant moments obtained from the data 
oscillate rather irregularly as the rank of the moments increases.  
Those behavior are well described by the calculated results by the 
MNBD as well as by the NBD.  The results in $pp$ and $\bar{p}p$ 
collisions are much different from those in $e^+e^-$ collisions, 
where the data is described much better by the MNBD than by the NBD. 

The cumulant moments of negatively charged particles obtained from 
the data in $pp$ or $\bar{p}p$ collisions does not show the regular 
oscillations contrary to those in $e^+e^-$ collisions.  
  
The characteristics of the cumulant moments in $hh$ and $e^+e^-$ 
collisions, discussed above,  are reduced mainly to the difference ofd
the hadronization between the two processes. 

\vspace{5mm}
{\bf Acknowledgements}

One of the authors  M. B. is partially supported by the Grant-in Aid 
for Scientific Research from the Ministry of Education, Science and 
Culture (No. 06640383).  N. S. thanks for the financial support by 
Matsusho Gakuen Junior College.

\newpage
%
\begin{flushleft}
{\large Table caption}
\end{flushleft}
\begin{itemize}

\item[Table 1] The parameters of the MNBD and the NBD used in the 
analysis of the cumulant moments in $pp$ collisions, and in $\bar
{p}p$ collisions.

\end{itemize}

\vspace{1cm}
\begin{flushleft}
{\large Figure captions}
\end{flushleft}
\begin{itemize}
\item[Fig. 1] The normalized cumulant moments of charged particles 
in 
$pp$ collisions.  The full circles are obtained from the data \cite
{brea84} at a) $\sqrt{s}=30.4$ GeV , b) $\sqrt{s}=44.5$ GeV, c) 
$\sqrt{s}=52.6$ GeV, and d) $\sqrt{s}=62.2$ GeV. The solid line is 
obtained from our calculations with the MNBD, and the dashed line is 
with the NBD.
\item[Fig. 2] The normalized cumulant moments of charged particles in 
$\bar{p}p$ collisions.  The full circles are obtained from the data 
\cite{anso87} from a) $\sqrt{s}=200$ GeV, b) $\sqrt{s}=546$ GeV, and 
c) $\sqrt{s}=900$ GeV. The dashed lines are obtained from our 
calculations.
\end{itemize}
\newpage
\begin{center}
\begin{tabular}{cccccc}
\hline
MNBD&$\sqrt{s}\ \ \rm[GeV]$&N&$\la n \ra$&$C_{2}$&$\chi^{2}$/NDF\\
\hline
    &30.4& 6 &4.214$\pm$0.062 & 1.2993$\pm$0.0034 &11.0/11\\
    &44.5& 6 &5.064$\pm$0.053 & 1.2775$\pm$0.0052 & 6.8/16\\
    &52.6& 7 &5.373$\pm$0.049 & 1.2880$\pm$0.0067 & 5.0/18\\
    &62.2&12 &5.821$\pm$0.061 & 1.2660$\pm$0.0077 &23.3/17\\
    &200 & 4 &10.69$\pm$0.13  & 1.2633$\pm$0.0113 & 7.7/28\\
    &546 & 4 &14.65$\pm$0.09  & 1.2746$\pm$0.0040 &67.0/44\\
    &900 & 3 &17.90$\pm$0.18  & 1.2955$\pm$0.0082 &57.7/52\\
\hline
\hline
 NBD&$\sqrt{s}\ \ \rm[GeV]$&&$\la n \ra$&$k$&$\chi^{2}$/NDF\\
\hline
    &30.4&   &4.370$\pm$0.059 & 21.44$\pm$3.76    &27.5/12\\
    &44.5&   &5.086$\pm$0.051 & 13.73$\pm$1.51    &14.7/17\\
    &52.6&   &5.384$\pm$0.049 &  9.98$\pm$0.65    & 5.5/19\\
    &62.2&   &5.823$\pm$0.062 & 10.77$\pm$0.79    &23.9/18\\
    &200 &   &10.56$\pm$0.13  &  5.66$\pm$0.36    & 4.3/29\\
    &546 &   &14.65$\pm$0.09  &  4.82$\pm$0.10    &89.6/45\\
    &900 &   &16.81$\pm$0.18  &  3.47$\pm$0.12    &73.7/53\\
\hline
\end{tabular}
\\
\vspace{1cm}
  Table 1 
\end{center}

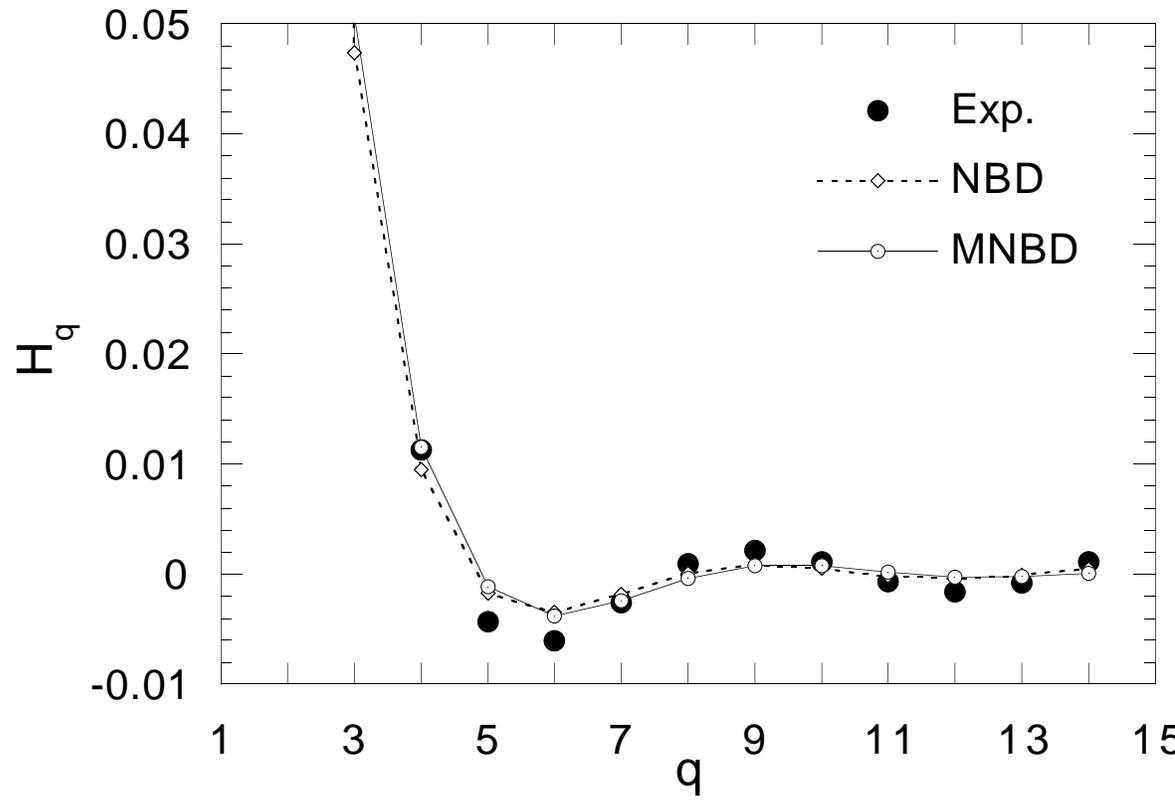
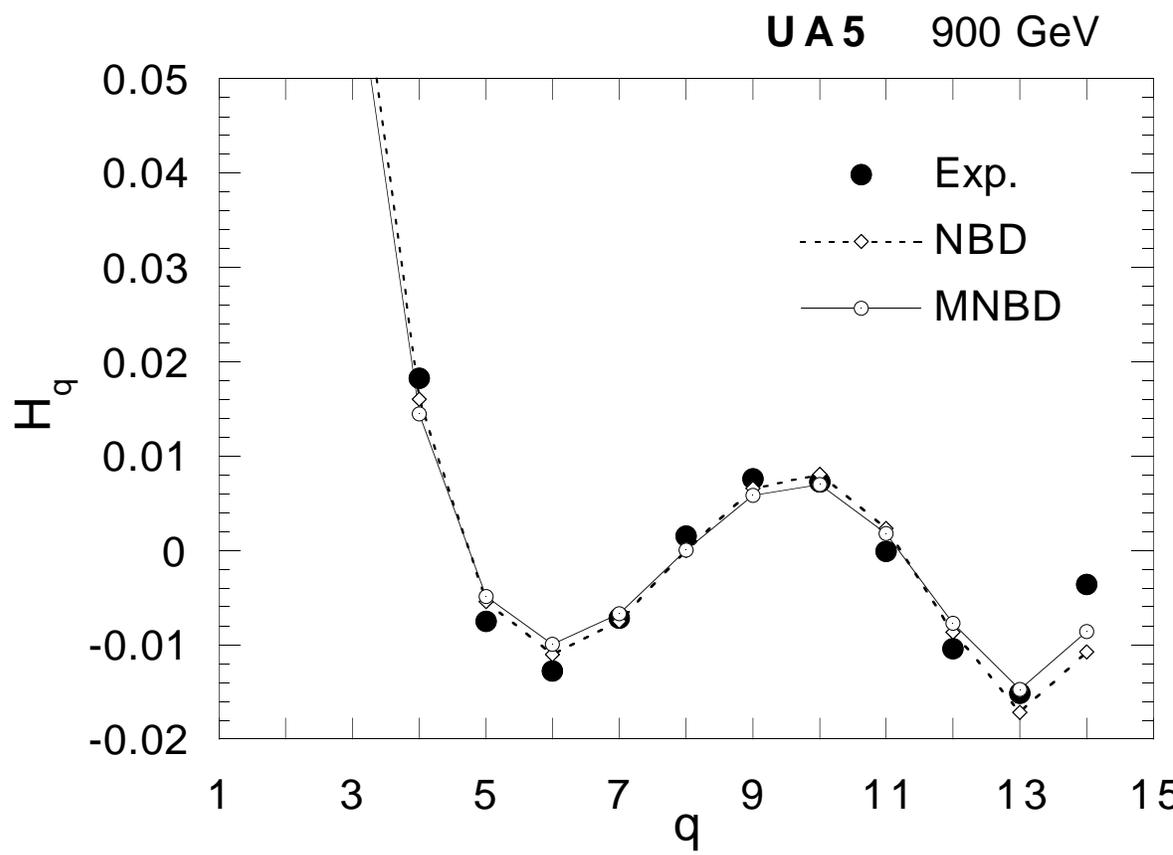
\end{document}